\documentclass[aps,prd]{revtex4}
\usepackage{graphics,psfrag}
\usepackage{graphicx,psfrag}
\usepackage{epic,eepic}
\usepackage{color} 
\usepackage{epsfig}
\usepackage{latexsym}
\usepackage{pstricks}
\newcommand{\be}{\begin{equation}}
\newcommand{\ee}{\end{equation}}
\newcommand{\bear}{\begin{eqnarray}}
\newcommand{\eear}{\end{eqnarray}}
\begin{document}

\title{Light bending in a Coulombic field}
\author{Jin Young Kim}
\email{jykim@kunsan.ac.kr}
\author{Taekoon Lee}
\email{tlee@kunsan.ac.kr}
\affiliation{Department of Physics, Kunsan National University, 
Kunsan 573-701, Korea}


\begin{abstract}
The nonlinear Euler-Heisenberg interaction bends light toward an
electric charge. The bending angle and trajectory of light in a
Coulombic field are computed in geometric optics.
\end{abstract}

 

\maketitle

The light bending by a massive object is one of the prominent features of the
general relativity and  is a useful tool in astrophysics through the
gravitational lensing. Though not directly related in physics, a question
may be raised whether there is an electrodynamic version of the bending:
that is, whether an electric charge can bend light toward, or outward of it.
Though it may most probably be very difficult to observe the effect 
we, nevertheless,
think it is interesting to address the question.

At classical level the linearity of the
electrodynamics precludes bending of light,
and therefore any bending must involve a nonlinear interaction from 
quantum corrections. The 
Euler-Heisenberg interaction that arises from the box diagram in
quantum electrodynamics can provide such a nonlinear interaction.

In this note we show that
an electric charge
bends light toward it through the Euler-Heisenberg interaction,
and compute the bending angle and trajectory of light 
in a Coulombic field. 
The bending of light by Euler-Heisenberg interaction is not new and
has been investigated by several authors, particularly on astronomical objects.
For instance, Denisov, et al. \cite{denisov} studied light bending
in the dipole magnetic field of a neutron star 
and De Lorenci, et al. \cite{lorenci1} studied the light bending by
a charged black hole. But none of the studies specifically addresses
our question above, and in addition  we develop a
 simple geometric way of
computing the bending angle and trajectory
 based on the Snell's law.

The box diagram of  quantum electrodynamics gives rise to 
a low energy effective Lagrangian of Euler-Heisenberg \cite{he,schwinger}
\bear
{\cal L}= -\frac{1}{4} F_{\mu\nu}F^{\mu\nu} +
\frac{\alpha^2\hbar^3}{90m^4c}\left[( F_{\mu\nu}F^{\mu\nu})^2+\frac{7}{4}
( F_{\mu\nu}\tilde F^{\mu\nu})^2\right]\,.
\label{lagrangian}
\eear  
In the presence of a background electric field the nonlinear
interaction modifies the dispersion relation for the electromagnetic
wave  and results in a modified speed of light that
 reads \cite{birula,adler,heyl,gies1,gies2,lorenci,holten}
\bear
\frac{v}{c}=1-\frac{a\alpha^2\hbar^3}{45m^4c^5}
({\bf u\times E})^2\,,
\eear
where $\bf u$ denotes the unit vector in the
direction of propagation. There is birefringence effect of the
light speed depending on
the photon polarization; The constant $a$ is either 14 for the normal mode
or 8 for the parallel mode, where the photon
polarization is, respectively, perpendicular to or
 parallel to the plane spanned by
${\bf u}$ and ${\bf E}$.
Because the speed of light depends on the field strength the light
bends in presence of a nonuniform field. The bending can be studied
in geometric optics by noting that the index of refraction of the
background field is given in leading order by 
\bear
n=\frac{c}{v}= 1+\frac{a\alpha^2\hbar^3}{45m^4c^5}
({\bf u\times E})^2\,.
\eear

\begin{figure}[htbp]
\begin{center}
\input{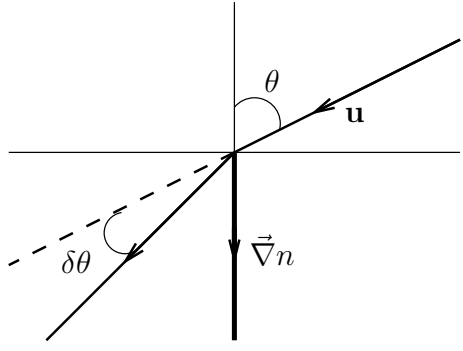}
\end{center}
\caption{Schematic of bending due to nonuniform refractive index.}
\label{fig1}
\end{figure}

The infinitesimal  bending of photon trajectory 
over $\delta \vec{r}$ can be obtained from
the Snell's law as
\bear
\delta \theta &=&\tan\theta\frac{\delta n}{n}\nonumber\\
              &=&\frac{1}{n}|\vec{\nabla}n\times \delta\vec{r}|\,,
\label{deltheta}
\eear
where $\theta$ denotes the angle between ${\bf u}$ 
and $\vec{\nabla}n$ (see Fig.\ref{fig1}).

For a Coulombic field of charge $q$
\bear
{\bf E}=\frac{q}{4\pi r^2} \hat{r}
\eear
and for a photon trajectory $y(x)$ on the $x$-$y$ plane
the refractive index can be written as
\bear
n=1+\frac{a\alpha^2\hbar^3 q^2}{720\pi^2 m^4c^5}\frac{(y-xy')^2}{r^6(1+y'^2)}\,.
\eear
Noticing that $\delta \theta$ in (\ref{deltheta})
can be as well thought as the change in angle between ${\bf u}$ and
the $x$-axis 
the equation for photon trajectory can be easily derived as
\bear
y''=\frac{1}{n}(1+y'^2)(\eta_2-\eta_1 y')
\label{trajectory}
\eear
where $\eta_{1,2}$, respectively, denotes the first two components of 
$\vec{\nabla}n$.

Now for an incoming photon with impact parameter $b$
the initial condition reads
\bear
y(-\infty)=b\,, \quad y'(-\infty)=0\,,
\label{inicond}
\eear
and the trajectory equation (\ref{trajectory}) 
becomes to the leading order
\bear
y''=\eta_2
\label{trajectory1}
\eear
with $\eta_2$ at leading order given by
\bear
\eta_2=\frac{a\alpha^2 q^2\lambda_e^4}
{360\pi^2\hbar c}(\frac{y}{r^6}-\frac{3y^3}{r^8})\,.
\eear

\begin{figure}[htbp]
\psfrag{U}[]{$U(\chi)$}
\psfrag{V}[]{$V(\chi)$}
\psfrag{x}[]{$\chi$}
\begin{center}
\includegraphics[width=0.90\textwidth,clip=true]{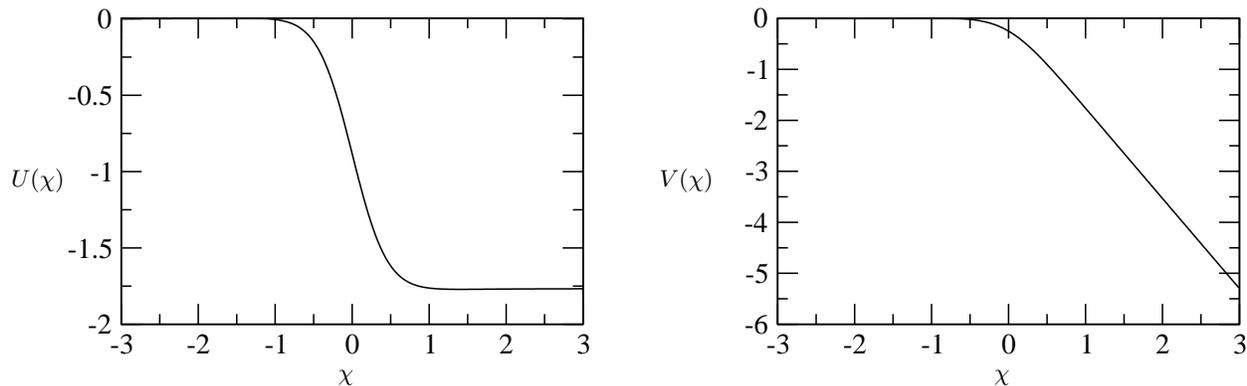}
\end{center}
\caption{Profile functions for bending angle and 
trajectory in Coulombic field.}
\label{fig2}
\end{figure}

The equation (\ref{trajectory1}) can be easily 
integrated by putting $y\!\!=\!\!b$ in $\eta_2$
for leading order solution
to obtain 
\bear
y'(x)&=&\frac{a\alpha^2 q^2}{360\pi^2\hbar c}
\left(\frac{\lambda_e}{b}\right)^4 U(\chi)\,,\nonumber\\
y(x)&=&b\left[1+\frac{a\alpha^2 q^2}{360\pi^2\hbar c}
\left(\frac{\lambda_e}{b}\right)^4 V(\chi)\right]\,,
\eear
where
\bear
U(\chi)&=&-
\frac{9\pi}{32}-
\frac{\chi(23+24 \chi^2+9 \chi^4)+9(1+\chi^2)^3
\arctan(\chi)}{16(1+\chi^2)^3}\,,\nonumber \\
V(\chi)&=&-\frac{9}{16}-
\frac{9\pi}{32}\chi +\frac{1}{8(1+\chi^2)^2}+
\frac{3}{16(1+\chi^2)}-\frac{9}{16} 
\chi \arctan (\chi)\,,
\eear
and $\chi=x/b$, and $\lambda_e=\hbar/mc$ denotes 
the Compton length of the electron.
The plots of the profile functions $U$ and $V$ for the bending
angle and the trajectory show the
bending occurs mostly over the 
region of $|x|\leq b$, and is attractive 
toward the charge (see Fig.\ref{fig2}).

The total bending angle $\theta$ is given by $|y'(\infty)|$ and reads
\bear
\theta =|y'(\infty)|
              =\frac{a\alpha^2 q^2}{640\pi\hbar c}
\left(\frac{\lambda_e}{b}\right)^4\,.
\label{bendingangle}
\eear

Let us now briefly comment on the experimental implication
of the bending in the Coulombic field by a nucleus of charge $Ze$.
For a nucleus of large $Z$  the effect of the bending
by the strong electric field may be dominant over the perturbative
quantum electrodynamics backgrounds such as the Compton scattering.  
To see this we may compare the cross section of the bending to that
of the Compton scattering.
Using the classical theory of scattering 
\cite{goldstein} and putting $q=Ze$ in (\ref{bendingangle})
we get the cross section  at small $\theta$
\bear
\frac{d\sigma}{d\Omega} 
&\approx&\frac{b}{\theta}\left|\frac{db}{d\theta}\right| \nonumber\\
                        &=&\frac{1}{4}\sqrt{\frac{a Z^2\alpha^3}{160}}
                        \frac{\lambda_e^2}{\theta^{\frac{5}{2}}}\,,
\eear
which is characterized by the 
Compton length of the electron (Note that $\alpha=e^2/4\pi\hbar c$).
Interestingly, the dependence of the cross section on $\alpha$ is 
nonanalytic. Its origin is dimensional;
The only dimensional parameter in the Lagrangian 
(\ref{lagrangian}) is the coefficient ($\sim \alpha^2/m^4$)
of the Euler-Heisenberg term and because the bending of light 
is achromatic the cross section must be proportional to the
square root of the coefficient. This is why the cross section is
proportional to $\sqrt{\alpha^3/m^4}$, 
with the one extra power of $\alpha$ 
coming from the Coulombic field.

Now compare the result to 
the   cross section for Compton scattering
$\sim Z^4\alpha^2\lambda_{\rm nucl.}^2$ that is set by the much 
smaller nucleus Compton length $\lambda_{\rm nucl.}$, which shows
the bending effect can easily dominate 
the Compton scattering, hence may be observable.
Nevertheless, the observation may require extreme precision.
The bending angle (\ref{bendingangle}) is valid provided the effective
Lagrangian of Euler-Heisenberg  as well as the geometric optics of
the light ray are applicable. The first requires the wavelength $\lambda$
of the photon be larger than the electron Compton length, and for
the geometric optics applicable the photon in the ray must do multiple
scattering via the box diagram in the Coulombic background; This
requires the impact parameter $b$, over which distance the bending
occurs mostly, be larger than both the electron Compton length and the
photon wavelength. Thus $\lambda_e<\lambda<b$ is required for the
bending angle (\ref{bendingangle}) to be valid. 
This condition now demands the bending angle be very small. As an example,
for $Z=100, b=10\lambda_e$ we get the bending angle  
$\theta=3.4\times 10^{-8}$ radian for 
an x-ray of wavelength $5\lambda_e$.

\begin{acknowledgments}
We are grateful to M. Park for useful discussions. 
\end{acknowledgments}

\bibliographystyle{apsrev}
\bibliography{lightbending}

\end{document}